\documentclass[aps,prd,twocolumn,superscriptaddress,nofootinbib]{revtex4-1}

\usepackage{latexsym}
\usepackage{amsmath}
\usepackage{amssymb}
\usepackage{amsfonts}
\usepackage{bm}
\RequirePackage{lineno}

\usepackage{color}
\definecolor{purple}{rgb}{0.5,0,0.5}
\definecolor{blue}{rgb}{0.0,0,0.9}
\definecolor{prdblue}{rgb}{0.133,0.118,0.498}
\usepackage[colorlinks=true, pdfstartview=FitV, linkcolor=prdblue, citecolor= prdblue, urlcolor=prdblue]{hyperref}
\newcommand{\BESIIIorcid}[1]{\href{https://orcid.org/#1}{\hspace*{0.1em}\raisebox{-0.45ex}{\includegraphics[width=1em]{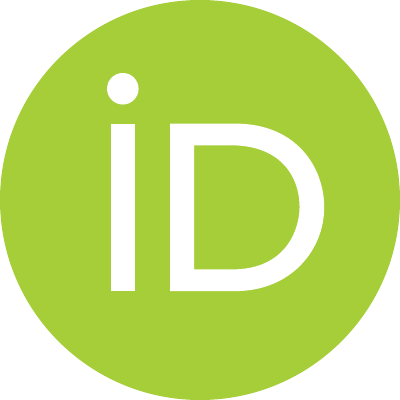}}}}

\usepackage{supertabular} 
\usepackage{placeins}
\usepackage{epsfig}
\usepackage{graphicx}
\usepackage{hyperref}

\hypersetup{
  breaklinks=true,
  colorlinks=true,
  linkcolor=blue,
  anchorcolor=blue,
  citecolor=blue,
  filecolor=blue,
  urlcolor=blue
}
\hypersetup{
  bookmarksnumbered=true,
  bookmarkstype=toc,
  linktocpage=true
}

\begin{document}

\modulolinenumbers[2]

\setlength{\oddsidemargin}{-0.5cm} \addtolength{\topmargin}{15mm}

\title{\boldmath  Precise measurement of the form factors in $D^0\rightarrow K^*(892)^-\ell^+\nu_{\ell}$ and observation of $D^0\rightarrow K_2^*(1430)^-\ell^+\nu_{\ell}$}

\author{
  \small
M.~Ablikim$^{1}$\BESIIIorcid{0000-0002-3935-619X},
M.~N.~Achasov$^{4,b}$\BESIIIorcid{0000-0002-9400-8622},
P.~Adlarson$^{81}$\BESIIIorcid{0000-0001-6280-3851},
X.~C.~Ai$^{86}$\BESIIIorcid{0000-0003-3856-2415},
R.~Aliberti$^{39}$\BESIIIorcid{0000-0003-3500-4012},
A.~Amoroso$^{80A,80C}$\BESIIIorcid{0000-0002-3095-8610},
Q.~An$^{77,64,\dagger}$,
Y.~Bai$^{62}$\BESIIIorcid{0000-0001-6593-5665},
O.~Bakina$^{40}$\BESIIIorcid{0009-0005-0719-7461},
Y.~Ban$^{50,g}$\BESIIIorcid{0000-0002-1912-0374}
% ...
 \\
 \vspace{0.2cm}
 (BESIII Collaboration)\\
 \vspace{0.2cm} {\it
$^{1}$ Institute of High Energy Physics, Beijing 100049, People's Republic of China\\
% ...
$^{4}$ Budker Institute of Nuclear Physics SB RAS (BINP), Novosibirsk 630090, Russia\\
% ...
$^{39}$ Johannes Gutenberg University of Mainz, Johann-Joachim-Becher-Weg 45, D-55099 Mainz, Germany\\
$^{40}$ Joint Institute for Nuclear Research, 141980 Dubna, Moscow region, Russia\\
% ...
$^{50}$ Peking University, Beijing 100871, People's Republic of China\\
% ...
$^{62}$ Southeast University, Nanjing 211100, People's Republic of China\\
% ...
$^{64}$ State Key Laboratory of Particle Detection and Electronics, Beijing 100049, Hefei 230026, People's Republic of China\\
% ...
$^{77}$ University of Science and Technology of China, Hefei 230026, People's Republic of China\\
% ...
$^{80}$ University of Turin and INFN, (A)University of Turin, I-10125, Turin, Italy; (B)University of Eastern Piedmont, I-15121, Alessandria, Italy; (C)INFN, I-10125, Turin, Italy\\
$^{81}$ Uppsala University, Box 516, SE-75120 Uppsala, Sweden\\
% ...
$^{86}$ Zhengzhou University, Zhengzhou 450001, People's Republic of China\\
\vspace{0.2cm}
% ...
$^{b}$ Also at the Novosibirsk State University, Novosibirsk, 630090, Russia\\
% ...
$^{g}$ Also at State Key Laboratory of Nuclear Physics and Technology, Peking University, Beijing 100871, People's Republic of China\\
% ...
\vspace{0.4cm}
}
}
\begin{abstract}
We report a study of the semileptonic decays $D^0 \rightarrow \bar{K}^0\pi^-\ell^+\nu_{\ell}$ (where $\ell=e,~\mu$) based on a sample of $20.3~\mathrm{fb}^{-1}$ of $e^+e^-$ annihilation data collected at a center-of-mass energy of 3.773~GeV with the BESIII detector at the BEPCII collider. 
Based on an investigation of the decay dynamics in $D^0 \rightarrow \bar{K}^0\pi^-\ell^+\nu_{\ell}$ decays, a $\mathcal{D}-$wave component of $D^0\rightarrow K_2^*(1430)^-\ell^+\nu_{\ell}$ is observed for the first time with a statistical significance of $8.0\sigma$, in addition to the dominant $K^*(892)^-$ and $\mathcal{S}-$wave components. The $\mathcal{D}-$wave component is determined to account for $(0.092 \pm 0.028_{\rm stat} \pm 0.018_{\rm syst})\%$ of the total decay rate. 
The branching fractions of the dominant $K^*(892)^-$ components are measured as $\mathcal{B}(D^0\rightarrow K^{*}(892)^-e^+\nu_{e}) = (2.043 \pm 0.018_{\rm stat} \pm 0.012_{\rm syst})\%$ and  $\mathcal{B}(D^0\rightarrow K^{*}(892)^-\mu^+\nu_{\mu}) = (1.964 \pm 0.018_{\rm stat} \pm 0.012_{\rm syst})\%$, which are the most precise measurements to date and represent significant improvements over the previous world averages. The hadronic form-factor parameters are measured to be $r_{V} = V(0)/A_1(0) = 1.444 \pm 0.026_{\rm stat} \pm 0.010_{\rm syst}$, $r_{2} = A_2(0)/A_1(0) = 0.752 \pm 0.020_{\rm stat} \pm 0.004_{\rm syst}$, and $A_1(0)=0.618\pm0.002_{\rm stat} \pm0.004_{\rm syst}$, where $V(0)$ is the vector form factor and $A_{1,2}(0)$ are the axial-vector form factors evaluated at $q^2=0$. This is the most precise determination of the form-factor parameters to date measured in a $D\rightarrow K^*(892)$ transition. In addition, we report the first model-independent measurement of the $\mathcal{S}-$wave phase shift in the hadronic $\bar{K}^0\pi^-$ system.
\end{abstract}

\pacs{13.30.Ce, 14.40.Lb, 14.65.Dw}

\maketitle

Semileptonic (SL) decays of $D$ mesons are an excellent laboratory to examine the interplay between the weak and strong interactions in hadron decays~\cite{physrept494,RevModPhys67_893}.
Indeed, the partial decay rate is related to the product of a hadronic form factor (FF) describing the effects of the strong interaction in the initial and final hadrons, and the Cabibbo-Kobayashi-Maskawa (CKM) matrix~\cite{prl10_531} element $|V_{cs(d)}|$ parametrizing the mixing of quarks in the weak interaction. 
Due to $|V_{cs(d)}|$ being tightly constrained by CKM unitarity, studies of $D$ meson SL decays provide an ideal laboratory to extract the hadronic FFs. 
In recent years, several hadronic FF measurements have been carried out in $D\rightarrow V\ell^+\nu_{\ell}$ decays~\cite{prl122_061801,prd78_051101,prd74_052001,prd81_112001,prd83_072001,prd94_032001,plb607_67,prd99_011103,D0Kspiev,2403.10877,prl110_131802,prd92_071101,2504.10867}, where $V$ and $\ell$ refer to a vector meson and a lepton, respectively. Throughout this Letter, charge-conjugate modes are implied unless explicitly noted. 
Furthermore, the SL decay $D^0\rightarrow K^{*}(892)^-\ell^+\nu_{\ell}$ also plays a crucial role in testing the standard model (SM), as it provides the most direct way to experimentally extract the value of $|V_{cs}|$. However, such an extraction requires precise theoretical knowledge of the hadronic FFs taking into account non-perturbative quantum chromodynamic (QCD) effects.

Although the study of the $D\rightarrow K^*(892)$ transition is very challenging, many theoretical calculations are performed with a variety of non-perturbative approaches~\cite{Lattice,IJMPA21_6125,PRR2,2505.01329,PRD62_014006,PRD96_016017,PRD67_014024,PRD72_034029,JPG39_025005,EPJC77_587,PRD89_034013,FrontPhys14_64401,prd101_013004,PRD92_054038,prd109_026008}.
In these calculations, one vector FF $V(q^2)$, and two axial-vector FFs $A_{1,2}(q^2)$, are introduced to describe the effects of the strong interactions.
However, the FF parameters of $r_V$ and $r_2$ in $D\rightarrow K^*(892)$ transition predicted by these  approaches~\cite{Lattice,IJMPA21_6125,PRR2,2505.01329,PRD62_014006,PRD96_016017,PRD67_014024,PRD72_034029,JPG39_025005,EPJC77_587,PRD89_034013,FrontPhys14_64401,prd101_013004,PRD92_054038,prd109_026008} differ significantly, and vary in the ranges from $1.36-1.78$ and $0.50-0.93$, respectively. 
Currently, the precision of $r_V$ and $r_2$ measurements is still insufficient to test these theoretical calculations~\cite{pdg24}, but improved measurements are now possible.  
These data can also motivate improved Lattice QCD calculations~\cite{Lattice}, and will also benefit the extraction of $|V_{cs}|$ via the $D\rightarrow K^*(892)$ transition.

The SL decays of $D$ mesons offer an excellent opportunity to test lepton flavor universality (LFU) of the SM~\cite{ARNPS,NSR}. In Refs.~\cite{PRD91_094009,CPC45_063107,PRD96_016017,PRR2}, it is suggested that observable LFU violation effects may occur in SL decays via $c\rightarrow s\ell^+\nu_{\ell}$. In particular, the multiple polarization states of vector mesons in $D\rightarrow V\ell^+\nu_{\ell}$ decay provide more physical information,
allowing a more detailed search for potential new physics effects~\cite{PRD91_094009}.
In recent years, the branching fraction (BF) of $D^0\rightarrow K^*(892)^-\ell^+\nu_{\ell}$ and the ratio $R^{\mu/e}_{K^*(892)}=\frac{\mathcal{B}(D^0\rightarrow K^*(892)^-\mu^+\nu_{\mu})}{\mathcal{B}(D^0\rightarrow K^*(892)^-e^+\nu_{e})}$ have been calculated using various theoretical approaches~\cite{IJMPA21_6125,PRR2,2505.01329,PRD62_014006,PRD96_016017,PRD67_014024,PRD72_034029,JPG39_025005,EPJC77_587,PRD89_034013,FrontPhys14_64401,prd101_013004,PRD92_054038}, leading to differing expectations. The predicted BFs for $e$ and $\mu$ modes varies from $(1.76-3.25)\%$ and $(1.66-3.09)\%$~\cite{PRD62_014006,PRD96_016017,PRD72_034029,IJMPA21_6125,JPG39_025005,EPJC77_587,FrontPhys14_64401,prd101_013004,PRD92_054038,PRR2,2505.01329}, and $R^{\mu/e}_{K^*(892)}$ is given as $0.92-0.95$ in Refs.~\cite{IJMPA21_6125,PRD96_016017,EPJC77_587,FrontPhys14_64401,PRD89_034013,prd101_013004,PRD92_054038} and $0.99$ in Ref.~\cite{PRR2}. 

The SL decay $D^0\rightarrow \bar{K}^0\pi^-\ell^+\nu_{\ell}$ also provides information on the interactions in the hadronic $K\pi$ system, and an opportunity to study the excited states of strange mesons $K_J^*$, where $J$ denotes spin. In comparison to the $K^-\pi^+$ system reported in Refs.~\cite{prd83_072001,prd94_032001,NPB133_490,NPB296_493}, a model-independent determination on the $\mathcal{S}-$wave phase ($\delta_S$) is still unavailable for the $\bar{K}^0\pi^-$ system. Comparisons of $\delta_S$ between the $K^-\pi^+$ and $\bar{K}^0\pi^-$ systems may also be made.  
Furthermore, the BFs of $D$ decays into a tensor strange meson, $D^0\rightarrow K_2^*(1430)^-\ell^+\nu_{\ell}$, were calculated based on the framework of the SU(3) flavor symmetry approach~\cite{EPJC84_1110} and the relativistic quark model (RQM)~\cite{prd111_093001}, with the predicted BFs ranges $(1.26-1.36)\times10^{-5}$ for the $e$ mode and $(0.81-0.91) \times 10^{-5}$ for the $\mu$ mode. The SL decay $D\rightarrow K_2^*(1430)\ell^+\nu_{\ell}$ were previously investigated in Refs.~\cite{prd83_072001,prd94_032001,D0Kspiev,2403.10877,2504.10867}. However, no experimental evidence has been found to date~\cite{pdg24}. 

In this Letter, we report a combined study of the dynamics in the SL decays $D^0\rightarrow \bar{K}^0\pi^-e^+\nu_{e}$ and $D^0\rightarrow \bar{K}^0\pi^-\mu^+\nu_{\mu}$.
We also report the most precise measurements of the BFs and FF parameters $r_V$ and $r_2$ in $D^0\rightarrow K^*(892)^-\ell^+\nu_{\ell}$ decays, and the first observation of the $D\rightarrow K_2^*(1430)\ell^+\nu_{\ell}$ transition, as well as the first model-independent measurement on the $\mathcal{S}-$wave phase in $\bar{K}^0\pi^-$ system.
These measurements are performed using a data sample corresponding to an integrated luminosity of $20.3~\mathrm{fb}^{-1}$~\cite{lum} recorded at the center-of-mass energy $\sqrt{s}=3.773$~GeV with the BEPCII $e^+e^-$ collider and collected by the BESIII detector~\cite{Ablikim:2009aa}.

Simulated data samples produced with a {\sc geant4}-based Monte Carlo (MC) package~\cite{geant4}, which includes the geometric description of the BESIII detector and the detector response, are used to determine detection efficiencies and to estimate background contributions. The simulation
models the beam energy spread and initial state radiation (ISR) in the $e^+e^-$ annihilations
with the {\sc kkmc} generator~\cite{kkmc}.
The inclusive MC sample is forty times larger than the corresponding data sample, and includes the production of $D\bar{D}$ pairs, the non-$D\bar{D}$ decays of the $\psi(3770)$, the ISR
production of the $J/\psi$ and $\psi(3686)$ states, and the continuum processes incorporated in {\sc kkmc}~\cite{kkmc}. 
All particle decays are modeled with {\sc evtgen}~\cite{nima462_152} using branching fractions either taken from
the Particle Data Group~\cite{pdg24}, when available, or otherwise estimated with {\sc lundcharm}~\cite{lundcharm}.
Final state radiation (FSR) from charged final state particles is incorporated using the
{\sc photos} package~\cite{plb303_163}. The generation of the signal $D^0\rightarrow \bar{K}^0\pi^-\ell^+\nu_{\ell}$ incorporates knowledge of the FFs obtained in this work.

Our analysis makes use of so called ``single-tag'' (ST) and ``double-tag'' (DT) samples of $D$ decays.
The ST samples are $\bar{D}^0$ events reconstructed from one of hadronic decays as listed in Table~\ref{tab:numST}, while the DT samples are events with a ST and a $D^0$ meson reconstructed as $D^0\rightarrow \bar{K}^0\pi^-\ell^+\nu_{\ell}$.  
The BF for the SL decay is given by~\cite{D0Kspiev}
\begin{equation}
  \mathcal{B}_{\rm SL} \,=\,
  \frac{N_{\rm DT}}{\sum_i N^{i}_{\rm ST} \,
      \left(\epsilon^i_{\rm DT}/\epsilon^i_{\rm ST}\right)} \,=\,
  \frac{N_{\rm DT}}{N_{\rm ST} \, \epsilon_{\rm SL}}, \label{eq:branch}
\end{equation}
where $N_{\rm DT(ST)}$ is the total yield of DT(ST) events, $\epsilon_{\rm SL}=(\sum_i N^{i}_{\rm ST} \, \epsilon^i_{\rm DT}/\epsilon^i_{\rm ST})/\sum_i N^{i}_{\rm ST}$ is the average efficiency of reconstructing the SL decay in a ST event,
weighted by the measured yields of ST modes in the data, and $\epsilon^i_{\rm ST}$ and $\epsilon^i_{\rm DT}$ are the efficiencies for finding the ST and the SL decay in the $i^{\rm-th}$ tag mode, respectively.

\begin{figure}[tp!]
\begin{center}
\includegraphics[width=\linewidth]{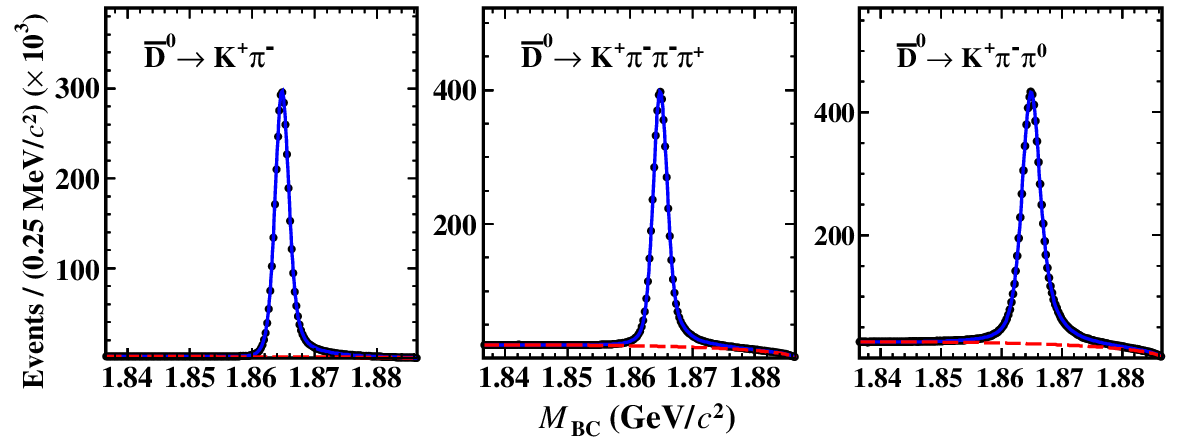}
\caption{(Color online)~The $M_{\rm BC}$ distributions of the three ST modes. The points are data, the solid blue curves are the total fits and the dashed red curves show the background shapes. }
\label{fig:tag_md0}
\end{center}

\end{figure}
\begin{table}[tp!]
\caption{ The selection requirements on $\Delta E$, the ST yields $N_{\rm ST}$ in data, the ST and DT efficiencies, $\epsilon_{\rm ST}$ and $\epsilon_{\rm DT}$, for each ST mode. For $\epsilon_{\rm DT}$ in the last column, the first number is for $D^0\rightarrow \bar{K}^0\pi^-e^+\nu_{e}$, and the second is for $D^0\rightarrow \bar{K}^0\pi^-\mu^+\nu_{\mu}$; the BF of $\bar{K}^0\rightarrow \pi^+\pi^-$ is not included. }
\begin{center}
\resizebox{!}{1.0cm}{
\begin{tabular}
{lcccc} \hline\hline ST mode  & $\Delta E$ (GeV)     &  $N^i_{\rm ST}$ ($\times 10^3$)   & $\epsilon^i_{\rm ST}$  (\%)    & $\epsilon^{i~e(\mu)}_{\rm DT}$  (\%)    \\
\hline $K^+\pi^-$                   & [$-$0.027, 0.027]          &  $3725.7\pm2.0$ & $65.10$     &   $18.96(13.10)$   \\
       $K^+\pi^-\pi^-\pi^+$       & [$-$0.026, 0.024]          &  $4987.5\pm2.5$ & $40.94$     &   $10.54(~6.54)$ \\
       $K^+\pi^-\pi^0$              & [$-$0.062, 0.049]          &  $7422.3\pm3.2$ & $35.60$     &  $~~9.87(~6.75)$   \\
\hline\hline
\end{tabular}
}
\label{tab:numST}
\end{center}
\end{table}

A detailed description of the selection criteria for $\pi^{\pm}$, $K^{\pm}$, $\gamma$, and $\pi^0$ candidates is given in Ref.~\cite{D0Kspiev}.
The ST $\bar{D}^0$ mesons are identified using the beam-constrained mass: 
\begin{equation}
M_{\rm BC} = \sqrt{(\sqrt{s}/2)^2-|\vec {p}_{\bar D^0}|^2},
\end{equation}
where $\vec {p}_{\bar D^0}$ is the momentum of the $\bar{D}^0$ candidate in the rest frame of the initial $e^+e^-$ system. 
A kinematic variable $\Delta E = E_{\bar{D}^0} -\sqrt{s}/2$ for
each candidate is used to improve the signal significance for ST $\bar{D}^0$ candidates, where $E_{\bar{D}^0}$ is the energy of the $\bar{D}^0$ candidate. 
The $\Delta E$ requirements for the ST modes are listed in Table~\ref{tab:numST}. 
The $M_{\rm BC}$ distributions for the three ST modes are shown in Fig.~\ref{fig:tag_md0}, along with the maximum likelihood fits. 
The signal functions are derived from the convolution of the MC-simulated signal shapes with a double-Gaussian smearing function to account for resolution difference between simulation and data; the parameters of the double-Gaussian function are floated. An ARGUS function~\cite{plb241_278} is used to describe the combinatorial background shape.
For each tag mode, the signal yield is obtained by integrating the signal
function over the $D^0$ region within $1.859<M_{\rm BC}<1.873$~GeV/$c^2$.
The yields and efficiencies for these ST modes are listed in Table~\ref{tab:numST}. The total ST yield summed over three hadronic modes is 
$N_{\rm ST}=(16135.4\pm4.6)\times 10^3$, where the uncertainty is statistical only.

\begin{figure}[tp!]
\begin{center}
   \includegraphics[width=\linewidth]{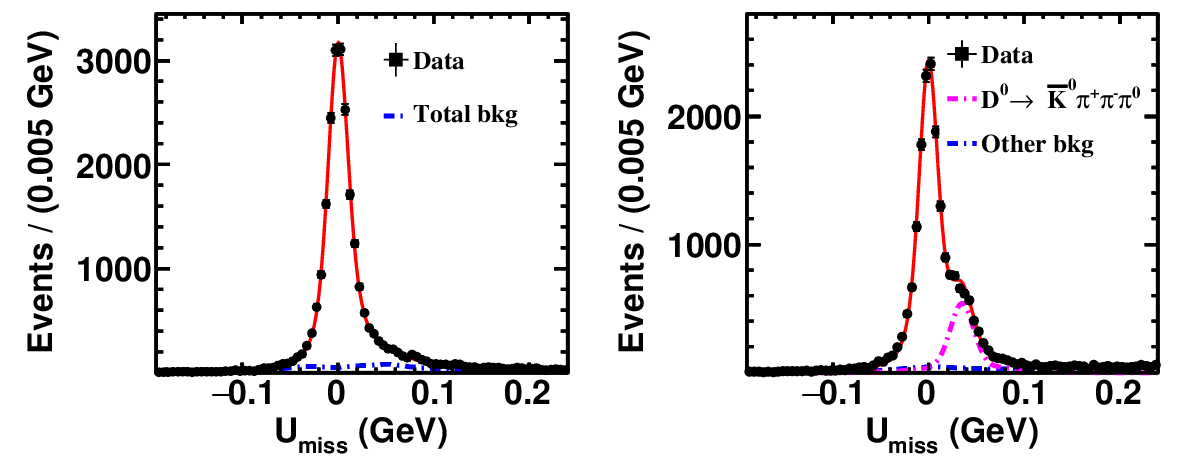}
   \caption{ (Color online)~Fits to $U_{\rm miss}$ distributions for the SL decays (left) $D^0\rightarrow \bar{K}^{0}\pi^- e^+\nu_{e}$ and (right) $D^0\rightarrow \bar{K}^{0}\pi^- \mu^+\nu_{\mu}$.  }
\label{fig:U}
\end{center}
\end{figure}

Candidates for the SL decay $D^0\rightarrow \bar{K}^0\pi^-e^+\nu_{e}$ and $D^0\rightarrow \bar{K}^0\pi^-\mu^+\nu_{\mu}$ are selected from the remaining tracks recoiling against the ST $\bar{D}^0$ mesons; the requirements given above for both $M_{\rm BC}$ and $\Delta E$ are applied to the ST. A detailed description of the selection criteria for the two SL decays is given in Ref.~\cite{D0Kspiev,2504.10867}.
As the neutrino is not detected, we employ the kinematic variable
$U_{\rm miss}=E_{\rm miss}-c|\vec{p}_{\rm miss}|$ to obtain information on the neutrino. Here, $E_{\rm miss}$ and $\vec{p}_{\rm miss}$ are the measured missing energy and momentum in the event, attributed to the unobserved neutrino; their definitions are given in Ref.~\cite{D0Kspiev,2504.10867}.
Figure~\ref{fig:U} shows the $U_{\rm miss}$ distributions of the accepted candidates for $D^0\rightarrow \bar{K}^{0}\pi^-e^+\nu_{e}$ and $D^0\rightarrow \bar{K}^{0}\pi^-\mu^+\nu_{\mu}$ in data.
To obtain the signal yields, unbinned maximum likelihood fits to the two $U_{\rm miss}$ distributions are performed. 
In each fit, the signal is described with a shape derived from simulated signal events convolved with a Gaussian function, where the mean and width of the Gaussian function are determined by the fit. The combinatorial background contribution is described using the shape obtained from the inclusive MC simulation. The peaking background from $D^0\rightarrow \bar{K}^0\pi^-\pi^+\pi^0$ to $D^0\rightarrow \bar{K}^{0}\pi^-\mu^+\nu_{\mu}$ decay is modeled using the MC-derived shape. The yields of $D^0\rightarrow \bar{K}^0\pi^-e^+\nu_{e}$ and $D^0\rightarrow \bar{K}^0\pi^-\mu^+\nu_{\mu}$ are $N^{e}_{\rm DT}=22171\pm183$ and $N^{\mu}_{\rm DT}=14212\pm138$, where the uncertainties are statistical only.

The DT efficiency $\epsilon^{i}$ from each tag mode is summarized in the last column of Table~\ref{tab:numST}. The average efficiencies for reconstructing SL decays, $D^0\rightarrow \bar{K}^0\pi^-e^+\nu_{e}$ and $D^0\rightarrow \bar{K}^0\pi^-\mu^+\nu_{\mu}$, are estimated to be $\epsilon^{e}_{\rm SL}=(28.34\pm0.01)\%$ and $\epsilon^{\mu}_{\rm SL}=(18.91\pm0.01)\%$, respectively, where the BF of $\bar{K}^0\rightarrow \pi^+\pi^-$ is not included. The DT efficiencies $\epsilon^{e}_{\rm SL}$ and $\epsilon^{\mu}_{\rm SL}$ are further corrected due to following issues. The difference due to positron and muon PID efficiencies between data and MC are estimated to be $(-2.8\pm0.1)\%$ and $(-3.5\pm0.2)\%$, respectively, studied by using the control samples of $e^+e^-\rightarrow \gamma e^+e^-$ and $e^+e^-\rightarrow \gamma \mu^+\mu^-$ events. The difference due to pion tracking efficiency between data and MC is estimated to be $(-0.4\pm0.1)\%$, evaluated using control samples of $D^0\rightarrow K^-\pi^+(\pi^0, \pi^+\pi^-)$, and $D^+\rightarrow K^-\pi^+\pi^+(\pi^0)$ events. In addition, the difference of the detection efficiency due to the $M_{\bar K^0\pi^-\mu^+(\pi^0)}$ requirement for the muon SL mode is estimated to be $(0.7\pm0.2)\%$, evaluated using a control sample of $D^+\rightarrow K^-\pi^+e^+\nu_{e}$ events. 
Hence, the DT efficiency for the $e$ SL mode is corrected by $-2.8\%$ and $-0.4\%$, giving $\epsilon^{e}_{\rm SL}=(27.44\pm0.01)\%$. The DT efficiency for the $\mu$ SL mode is corrected by $-3.5\%$, $-0.4\%$ and $+0.7\%$, giving $\epsilon^{\mu}_{\rm SL}=(18.30\pm0.01)\%$. Therefore, we measure the BFs of the two SL decays to be $\mathcal B({D^0\rightarrow \bar{K}^{0}\pi^-e^+\nu_{e}})=(1.447\pm0.012_{\rm stat})\%$ and $\mathcal B({D^0\rightarrow \bar{K}^{0}\pi^-\mu^+\nu_{\mu}})=(1.391\pm0.013_{\rm stat})\%$, respectively, where the uncertainties are statistical only.

Due to the DT technique, the BF measurement is insensitive to the systematic uncertainty in the ST efficiency.
The uncertainty on the positron tracking (PID) efficiency is estimated to be 0.1\%~(0.1\%) by studying a sample of $e^+e^-\rightarrow \gamma e^+e^-$ events.
The uncertainty on the muon tracking (PID) efficiency is estimated to be 0.1\%~(0.2\%) by studying a sample of $e^+e^-\rightarrow \gamma \mu^+\mu^-$ events.
The uncertainty due to the pion tracking (PID) efficiency is estimated to be 0.1\%~(0.1\%) using control samples selected from $D^0\rightarrow K^-\pi^+(\pi^0, \pi^+\pi^-)$, and $D^+\rightarrow K^-\pi^+\pi^+(\pi^0)$. The uncertainty from $K^0_S$ reconstruction is 0.2\%, determined with control samples selected from
$D^0 \rightarrow \bar{K}^0\pi^+\pi^-, \bar{K}^0\pi^+\pi^-\pi^0, \bar{K}^0\pi^0$, and  $D^+ \rightarrow \bar{K}^0\pi^+, \bar{K}^0\pi^+\pi^0, \bar{K}^0\pi^+\pi^+\pi^-$.
 The uncertainty associated with the $E_{\gamma\,{\rm \max}}$ requirement is estimated to be less than 0.1\% by analyzing DT $D^0\bar{D}^0$ events where $D^0$ mesons decay to hadronic final states of $D^0\rightarrow K^-\pi^+$, $K^-\pi^+\pi^0$, and $K^-\pi^+\pi^+\pi^-$. The uncertainties due to the $M_{\bar K^0\pi^-e^+}$ and $M_{\bar K^0\pi^-\mu^+(\pi^0)}$ requirements are estimated to be 0.2\% and 0.2\%, respectively, evaluated using a control sample from $D^+\rightarrow K^-\pi^+e^+\nu_{e}$. The uncertainty due to the modeling of the signal in simulated events is estimated to be 0.3\% by varying the input FF parameters determined in this work by $\pm 1\sigma$. The uncertainty associated with the fit to the $U_{\rm miss}$ distribution is estimated to be 0.3\% by varying the fitting ranges and the shapes which parametrize the signal and background candidates, where an asymmetric Gaussian function is used as an alternative signal function.
The uncertainty associated with the fit of the $M_{\rm BC}$ distributions used to determine $N_{\rm ST}$ is 0.3\% and is evaluated by varying the bin size, fit range and background distributions. Further systematic uncertainties are assigned due to the statistical precision of the simulation, 0.1\%, and the input BF of the decay $K^0_S\rightarrow \pi^+ \pi^-$, 0.1\%. 
All systematic uncertainties are summed in quadrature and the total systematic uncertainties on the measured BFs of $D^0\rightarrow \bar{K}^0\pi^-e^+\nu_{e}$ and $D^0\rightarrow \bar{K}^0\pi^-\mu^+\nu_{\mu}$, are determined to be 0.6\% each.
Finally, we obtain $\mathcal B({D^0\rightarrow \bar{K}^{0}\pi^-e^+\nu_{e}})=(1.447\pm0.012_{\rm stat}\pm0.009_{\rm syst})\%$ and $\mathcal B({D^0\rightarrow \bar{K}^{0}\pi^-\mu^+\nu_{\mu}})=(1.391\pm0.013_{\rm stat}\pm0.008_{\rm syst})\%$.

To investigate the decay dynamics of $D^0\rightarrow \bar{K}^{0}\pi^-\ell^+\nu_{\ell}$ decays, the signal candidates of $D^0\rightarrow \bar{K}^0\pi^-e^+\nu_{e}$ and $D^0\rightarrow \bar{K}^0\pi^-\mu^+\nu_{\mu}$ decays are analyzed simultaneously.
The differential decay rates of $D^0\rightarrow \bar{K}^{0}\pi^-\ell^+\nu_{\ell}$ can be expressed in terms of five kinematic variables:
the squared invariant mass of $\bar{K}^0\pi^-$ system ($m_{\bar{K}^0\pi^-}^2$), the squared transfer momentum of $\ell^+$ and $\nu_{\ell}$ ($q^2$),
the angle between $\pi^-$ and $D^0$ direction of flight in the $\bar{K}^0\pi^-$ rest frame ($\theta_{\bar{K}^0}$), the angle between neutrino $\nu$ and $D^0$ direction in the $\ell^+\nu_{\ell}$ rest frame ($\theta_{\ell}$), and the acoplanarity angle ($\chi$) between the decay planes of the $\bar{K}^0\pi^-$ and $\ell^+\nu_{\ell}$. 
\begin{linenomath*}
The differential decay rate is expressed as~\cite{prd46_5040,CPC47_063101}
\begin{eqnarray}
d^5\Gamma&=&\frac{G^2_F|V_{cs}|^2}{(4\pi)^6m^3_{D^0}}X \beta \, \beta_{\ell} \, \mathcal{I}(m_{\bar{K}^0\pi^-}^2, q^2, \theta_{\bar{K}^0}, \theta_{\ell}, \chi) \nonumber \\
 & & dm_{\bar{K}^0\pi^-}^2dq^2 \, d{\rm cos}\theta_{\bar{K}^0} \, d{\rm cos}\theta_{\ell} \, d\chi,
\label{eq:differential}
\end{eqnarray}
where $X=p_{\bar{K}^{0}\pi^-}m_{D^0}$, $\beta=2p^{*}/m_{\bar{K}^{0}\pi^-}$, $\beta_{\ell}=(1-m^2_{\ell}/q^2)$. Furthermore, $p_{\bar{K}^{0}\pi^-}$ is the momentum of the $\bar{K}^{0}\pi^-$ system in the $D^0$ rest frame, $p^*$ is the momentum of $\bar{K}^{0}$ in the $\bar{K}^{0}\pi^-$ rest frame, $m_{\ell}$ ($m_{D^0}$) is the known lepton ($D^0$) mass~\cite{pdg24}, and $G_F$ is the Fermi coupling constant. 
In addition to the $\mathcal{S}-$wave and $\mathcal{P}-$wave components incorporated in Refs.~\cite{D0Kspiev,2403.10877}, a new $\mathcal{D}$-wave component is taken into account in the decay density $\mathcal{I}$.
The detailed expression of $\mathcal{I}$ is documented in the supplementary materials~\cite{supple}. 
The related $\mathcal{D}$-wave FFs $\mathcal{F}_{12}$, $\mathcal{F}_{22}$ and $\mathcal{F}_{32}$ take the form of 
\begin{eqnarray}
\mathcal{F}_{12} &=& \frac{m_{D^0}p_{\bar{K}^0\pi^-}}{3} \left[-\frac{m^2_{D^0} \, p^2_{\bar{K}^0\pi^-}}{m_{D^0}+m_{\bar{K}^0\pi^-}} \, T_2(q^2) \right.\nonumber \\
                 &+&(m^2_{D^0}-m^2_{\bar{K}^0\pi^-}-q^2)(m_{D^0}+m_{\bar{K}^0\pi^-}) \nonumber \\
  & & \left.  T_1(q^2) \frac{}{}\right] \, \mathcal{A}^{\prime}(m), \nonumber \\
  \mathcal{F}_{22} &=& \sqrt{\frac{2}{3}}\, m_{D^0} \, m_{\bar{K}^0\pi^-} \, q \, p_{\bar{K}^0\pi^-}(m_{D^0}+m_{\bar{K}^0\pi^-}) \nonumber \\
                 & & T_1(q^2) \, \mathcal{A}^{\prime}(m), \nonumber \\
\mathcal{F}_{32} &=& \sqrt{\frac{2}{3}} \, \frac{2m^2_{D^0} \, m_{\bar{K}^0\pi^-} \, q \, p^2_{\bar{K}^0\pi^-}}{m_{D^0}+m_{\bar{K}^0\pi^-}} \, T_V(q^2) \, \mathcal{A}^{\prime}(m) \, . \nonumber
\end{eqnarray}
The $\mathcal{D}$-wave FFs are assumed to follow the simple pole model and contain one vector and two axial-vector FFs which are denoted as $T_V(q^2)$ and $T_{1,2}(q^2)$. 
The ratios of the FF parameters are then defined as $r_V^D=T_V(0)/T_1(0)$ and $r_2^D=T_2(0)/T_1(0)$ at $q^2=0$, where $r_V^D$ and $r_2^D$ are expected to be 1~\cite{prd94_032001,prd83_072001}. The amplitude $\mathcal{A}^{\prime}(m)$ takes a relativistic Breit-Wigner form with a mass-dependent width, 
$$ \mathcal{A}^{\prime}(m) = \frac{r_D \, m_{K_2^*(1430)} \, \Gamma^0_{K_2^*(1430)} \, F_2(m)}{m_{K_2^*(1430)}^2 - m^2 - i \, m_{K_2^*(1430)} \, \Gamma_{K_2^*(1430)}(m)},$$
where $r_D$ denotes the magnitude of the $\mathcal{D}-$wave amplitude, $F_2(m)=\left(\frac{p^*}{p_0^*}\right)^2\frac{B_2(p^*)}{B_2(p_0^*)}$ with $B_2(p^*)=1/\sqrt{(r_{BW}^2p^{*2}-3)^2+9r_{BW}^2p^{*2}}$, and $\Gamma_{K_2^*(1430)}(m)=\Gamma^0_{K_2^*(1430)}\left(\frac{p^*}{p^*_0}\right)\frac{m_{K_2^*(1430)}}{m_{\bar{K}^0\pi^-}} \, F^2_2(m)$ with the mass $m_{K_2^*(1430)}$ and width $\Gamma^0_{K_2^*(1430)}$ for $K_2^*(1430)$ fixed as in Ref.~\cite{pdg24}.
\end{linenomath*}

\begin{figure}[tp!]
\begin{center}
   \includegraphics[width=\linewidth]{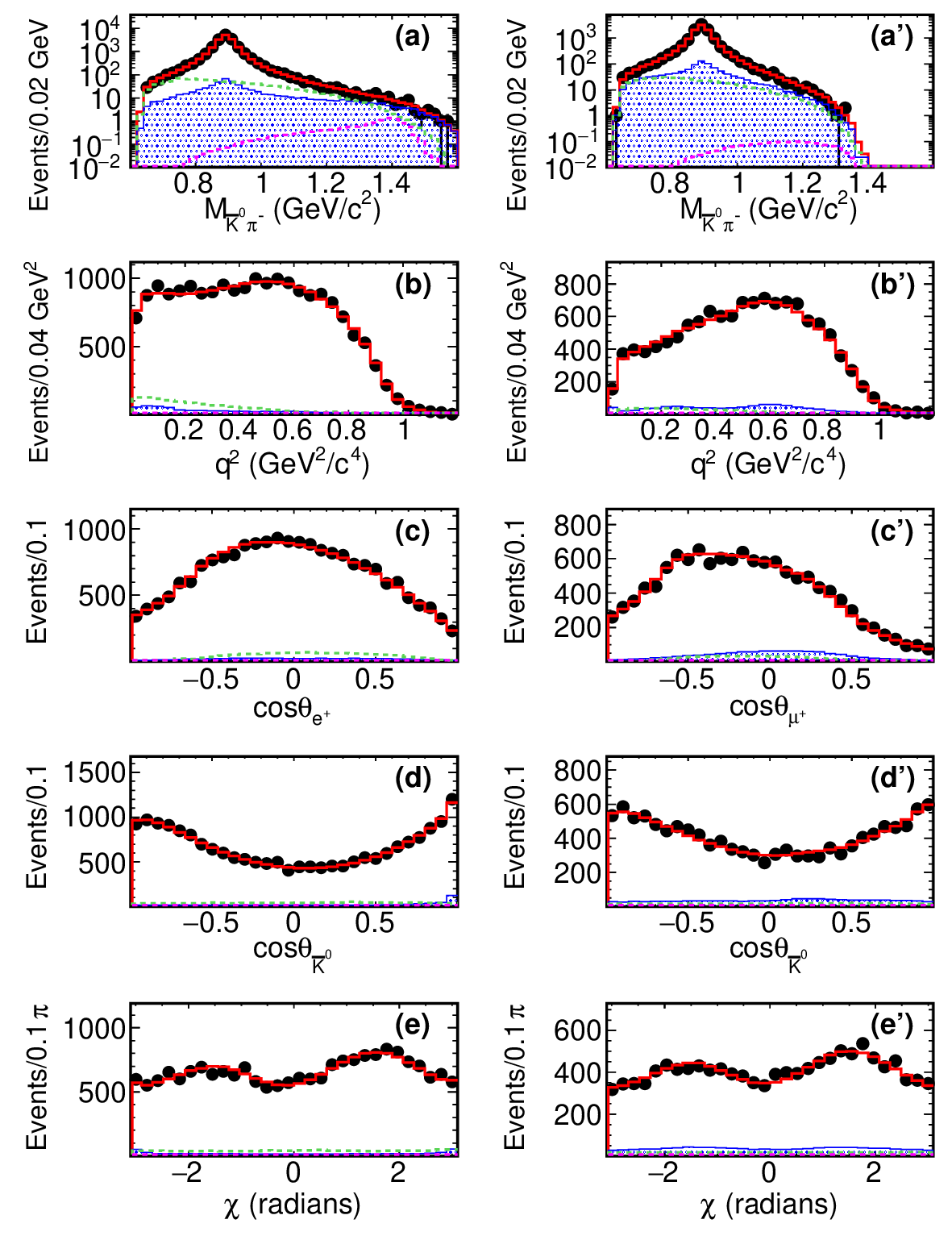}   
   \caption{ (Color online)~ Projections of the fits and data for (a, a$^{\prime}$) $M_{\bar{K}^0\pi^-}$, (b, b$^{\prime}$)  $q^2$, (c, c$^{\prime}$) $\cos\theta_{\ell^+}$, (d, d$^{\prime}$) $\cos\theta_{\bar{K}^0}$, and (e, e$^{\prime}$)  $\chi$ for the SL decay $D^0\rightarrow \bar{K}^0\pi^-\ell^+\nu_{\ell}$, where the dashed green and dashed pink curves show the contribution of the $\mathcal{S}-$wave and $\mathcal{D}-$wave components. The dots with error bars are data, the red histograms are the fit results, and the shaded histograms are the simulated background. }
\label{fig:FF}
\end{center}
\end{figure}

To extract the FF parameters, an unbinned five-dimensional maximum likelihood fit is performed to the distributions of $m_{\bar{K}^0\pi^-}$, $q^2$, $\cos\theta_{\bar{K}^0}$, $\cos\theta_{\ell}$, and $\chi$ observed in $D^0\rightarrow \bar{K}^{0}\pi^-\ell^+\nu_{\ell}$ decays. The signal candidates of $D^0\rightarrow \bar{K}^0\pi^-e^+\nu_{e}$ and $D^0\rightarrow \bar{K}^0\pi^-\mu^+\nu_{\mu}$ decays within $-0.05<U_{\rm miss}<0.05$~GeV and $-0.02<U_{\rm miss}<0.02$~GeV are selected.  
In the fit, the FF ratios $r_V$ and $r_2$, the mass $M_{K^{*}(892)^-}$ and width $\Gamma_{K^{*}(892)^-}$, the $\mathcal{D}$-wave amplitude $r_D$, and the parameters used to describe the $\mathcal{S}-$wave component including the relative intensity $r_S$, scattering length $a^{1/2}_{\rm S,BG}$, the effective range $b^{1/2}_{\rm S,BG}$, the dimensionless coefficient $r_S^{(1)}$ as defined in Ref.~\cite{supple}, are floating. 
The projected distributions of the fit onto the fitted variables are shown in Fig.~\ref{fig:FF}. The fit results are summarized in Table~\ref{tab:FitResults}. The fitted mass and width of the $K^*(892)^-$ are $892.7\pm0.2$~MeV/$c^2$ and $\Gamma_{K^*(892)^-}=45.6\pm0.4$~MeV, which are consistent with the world-average results~\cite{pdg24}. 
The statistical significance of the $\mathcal{D}$-wave component is determined to be $8.0\sigma$, calculated via $\sqrt{-2 \, \Delta {\rm ln}\mathcal{L}}$, where $\Delta {\rm ln}\mathcal{L}$ is the variation in ${\rm ln}\mathcal{L}$ of the likelihood fit with and without the $\mathcal{D}$-wave component included. A possible contribution from the other $\mathcal{P}$-wave component, $K^*(1410)$, is also tested. However, the statistical significance of incorporating an additional $K^*(1410)$ component is $2.7\sigma$ and it is thus neglected.

The fit fraction of each component can be determined by the ratio of the decay intensity of the specific component to that of the total intensity. The fractions of $\mathcal{S}$-wave, $\mathcal{P}$-wave ($K^{*}(892)^-$) and $\mathcal{D}$-wave ($K_2^{*}(1430)^-$) components are determined to be
$f_{S-{\rm wave}}=(5.81\pm0.19_{\rm stat})\%$, $f_{K^{*}(892)^-}=(94.12\pm0.19_{\rm stat})\%$, and $f_{K_2^{*}(1430)^-}=(0.092\pm0.028_{\rm stat})\%$, respectively, where the uncertainty propagation includes correlations among the underlying parameters.  The systematic uncertainties of the fit parameters and the fractions of $\mathcal{S}$-wave, $K^{*}(892)^-$ and $K_2^{*}(1430)^-$ components
are defined as the difference between the nominal fit and alternative fits with varied conditions.
The systematic uncertainties due to the requirements on $E_{\gamma\,{\rm \max}}$, $M_{\bar{K}^0\pi^-e^+}$ or $M_{\bar{K}^0\pi^-\mu^+(\pi^0)}$, $e/\mu$ requirements, background subtraction, tracking and particle identification (PID) efficiencies are evaluated in the same way as in Ref.~\cite{D0Kspiev,2504.10867}. The systematic uncertainty for each source is documented in the supplementary materials~\cite{supple}. The total absolute systematic uncertainty for each parameter is also listed in Table~\ref{tab:FitResults}. 

\begin{figure}[tp!]
\begin{center}
   \flushleft
   \begin{minipage}[t]{8cm}
   \includegraphics[width=\linewidth]{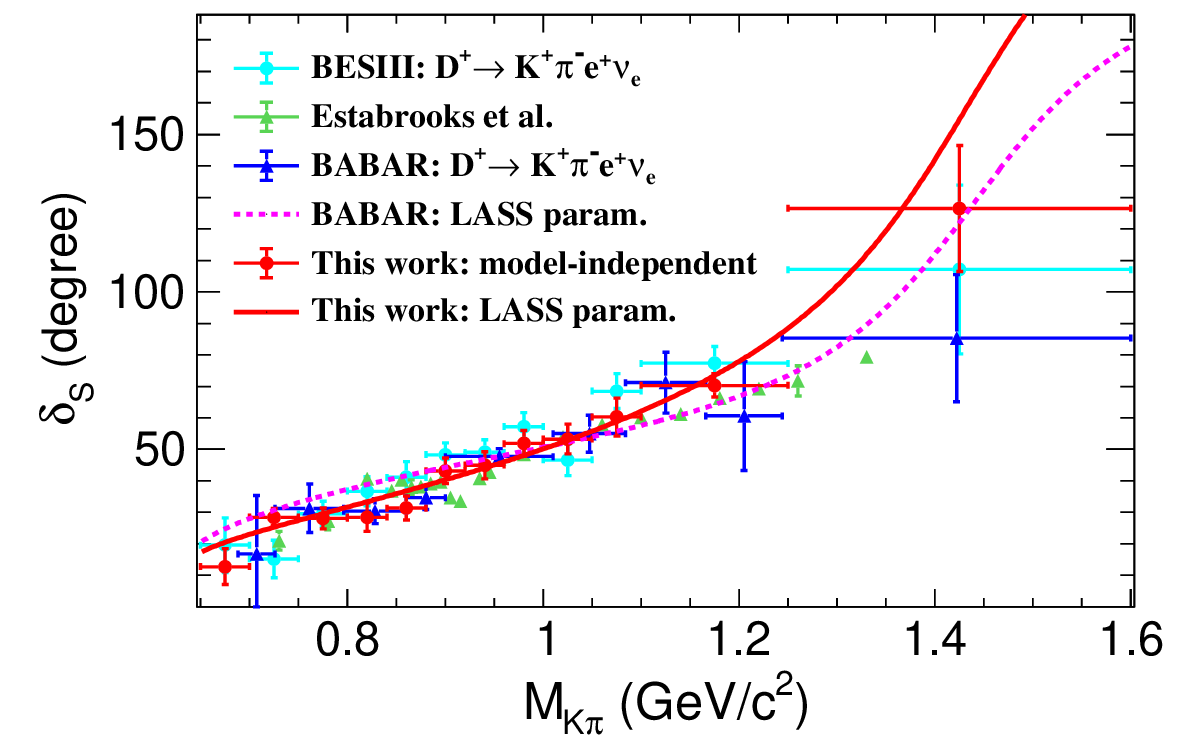}
   \end{minipage}    
   \caption{ (Color online)~Comparisons of the $\mathcal{S}-$wave phase shift $\delta_S$ measured in this work with others. The red dots with error bars show the model-independent measurements in this work; the solid red curve shows the result based on the LASS parameterization; the cyan dots, green dots and blue dots show measurements from BESIII~\cite{prd94_032001}, Refs.~\cite{NPB133_490,NPB296_493} and BaBar~\cite{prd83_072001}; The dashed pink curve show the result from BABAR based on the LASS parameterization~\cite{prd83_072001} with $a^{1/2}_{\rm S,BG}=1.95$ and $b^{1/2}_{\rm S,BG}=1.76$. }
\label{fig:Phase}
\end{center}
\end{figure}

We also perform a model-independent measurement on the phase shift of the $\mathcal{S}-$wave to validate the LASS parameterization~\cite{prd83_072001}. In this analysis, the $m_{\bar{K}^0\pi^-}$ spectrum from 0.6~GeV/$c^2$ to 1.6~GeV/$c^2$ is divided into 12 bins and the phase shift, $\delta_S$, in each $m_{\bar{K}^0\pi^-}$ interval is assumed to be constant. Then we re-perform the maximum likelihood fit as we do in previous studies, and the resultant values of $\delta_S$ in each of the 12 mass regions are obtained from the fit. The detailed results for $\delta_S$ versus mass documented in the supplementary materials~\cite{supple}. 
Comparisons of $\delta_S$ measured in this work with other results are shown in Fig.~\ref{fig:Phase}. 

\begin{table}
\begin{center}
\caption{The fit results, where the first uncertainty is statistical and the second is systematic. } \normalsize
\begin{tabular}
{lc} \hline\hline  Variable~~~~~~~~~~~~~~~~~~~~~~~     &   Value    \\ \hline
$r_V$                 &  $1.444\pm0.026\pm0.011$  \\
$r_2$                 &  $0.752\pm0.020\pm0.004$  \\
$r_S$                 &  $-13.21\pm0.49\pm0.36$ \\
$a^{1/2}_{\rm S,BG}$ (GeV/$c$)$^{-1}$ &  $1.24\pm0.11\pm0.08$  \\
$b^{1/2}_{\rm S,BG}$ (GeV/$c$)$^{-1}$ &  $-2.71\pm0.75\pm0.71$  \\
$r_S^{(1)}$        & $-0.05\pm0.04\pm0.03$ \\
$r_D$                & $11.52\pm2.00\pm1.16$ \\
$f_{K^*(892)^-}$          &  $(94.12\pm0.19\pm0.09)\%$  \\
$f_{S-{\rm wave}}$    &  $(5.81\pm0.19\pm0.09)\%$ \\
$f_{D-{\rm wave}}$    &  $(0.092\pm0.028\pm0.018)\%$ \\
\hline\hline 
\end{tabular}
\label{tab:FitResults}
\end{center}
\end{table}

\begin{figure}[tp!]
\begin{center}
   \begin{minipage}[t]{8cm}
   \includegraphics[width=\linewidth]{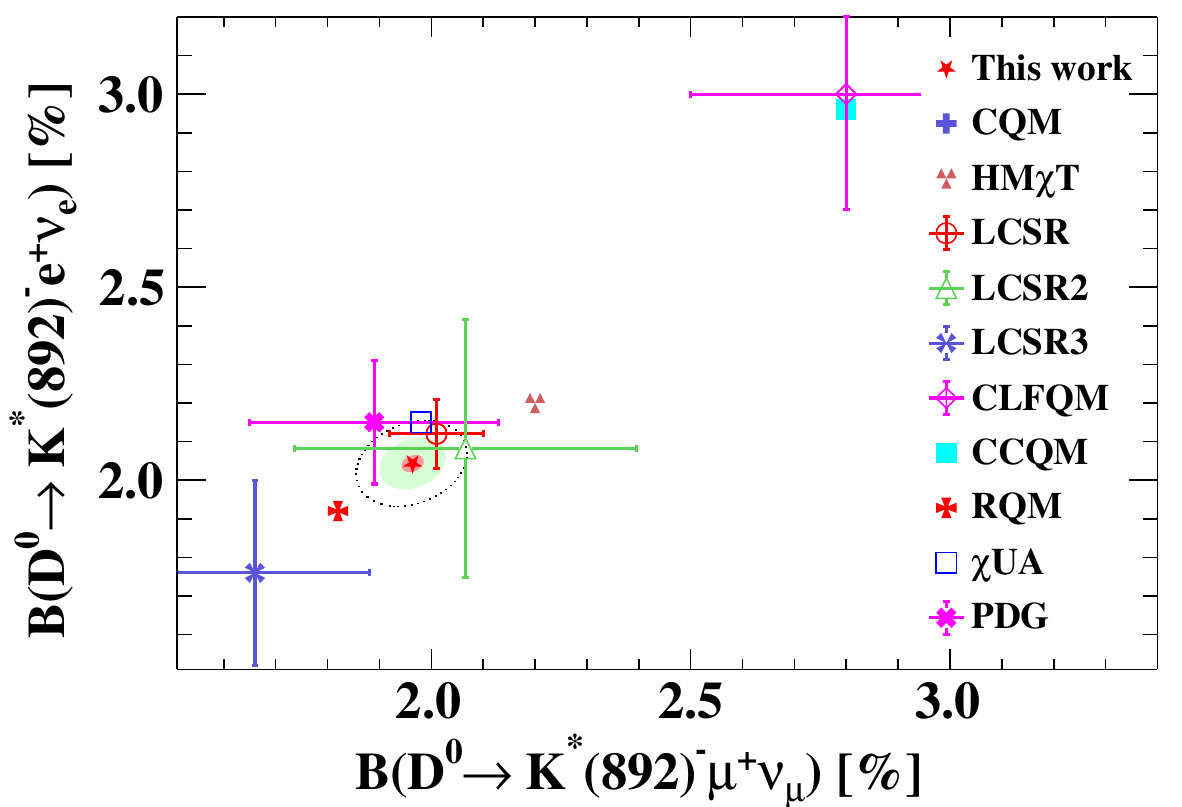}
   \end{minipage}    
   \caption{(Color online)~Comparisons of the measured $\mathcal{B}(D^0\rightarrow K^{*}(892)^-e^+\nu_{e})$ and $\mathcal{B}(D^0\rightarrow K^{*}(892)^-\mu^+\nu_{\mu})$ with various theoretical calculations from the CQM~\cite{PRD62_014006,PRD96_016017}, HM$\chi$T~\cite{PRD72_034029}, LCSR~\cite{IJMPA21_6125}, LCSR~2~\cite{PRR2}, LCSR~3~\cite{2505.01329}, CLFQM~\cite{JPG39_025005,EPJC77_587}, CCQM~\cite{FrontPhys14_64401}, RQM~\cite{prd101_013004}, $\chi$UA~\cite{PRD92_054038}, and PDG averaged results~\cite{pdg24}. The correlation coefficient between the measured BFs is $0.24$. The color coding for the measured results are the same as in Fig.~\ref{fig:cmpformfactor}.}   
\label{fig:cmpbf}
\end{center}
\end{figure}

In summary, with $20.3~\mathrm{fb}^{-1}$ of $e^+e^-$ annihilation data collected at $\sqrt{s}=3.773$ GeV by the BESIII detector, the absolute BFs of $D^0\rightarrow \bar{K}^0\pi^-\ell^+\nu_{\ell}$ decays are measured to be $\mathcal B({D^0\rightarrow \bar{K}^{0}\pi^-e^+\nu_{e}})=(1.447\pm0.012_{\rm stat}\pm0.009_{\rm syst})\%$ and $\mathcal B({D^0\rightarrow \bar{K}^{0}\pi^-\mu^+\nu_{\mu}})=(1.391\pm0.013_{\rm stat}\pm0.008_{\rm syst})\%$.
By analyzing the dynamics of the $D^0\rightarrow \bar{K}^0\pi^-\ell^+\nu_{\ell}$ decay, a non-trivial contribution of the $\mathcal{D}-$wave component $D^0\rightarrow K_2^*(1430)^-\ell^+\nu_{\ell}$ is observed for the first time with a statistical significance of $8.0\sigma$. The $e-\mu$ averaged $\mathcal{S}-$wave, $\mathcal{P}-$wave, and $\mathcal{D}-$wave components are determined to account for $(5.81\pm 0.19_{\rm stat} \pm 0.19_{\rm syst})\%$, $(94.12\pm0.19_{\rm stat}\pm0.09_{\rm syst})\%$, and $(0.092\pm0.028_{\rm stat}\pm0.018_{\rm syst})\%$ of the total decay rate.  
The BFs of the $\mathcal{S}$-wave components are $\mathcal{B}[D^0\rightarrow (\bar{K}^0\pi^-)_{S-{\rm wave}} \, e^+\nu_{e}] = (0.084 \pm 0.003_{\rm stat} \pm 0.001_{\rm syst})\%$ 
and $\mathcal{B}[D^0\rightarrow (\bar{K}^0\pi^-)_{S-{\rm wave}} \, \mu^+\nu_{\mu}] = (0.081 \pm 0.003_{\rm stat} \pm 0.001_{\rm syst})\%$.
The BFs of the dominant $K^{*}(892)^-$ components in the SL decays are measured to be $\mathcal{B}(D^0\rightarrow K^{*}(892)^-e^+\nu_{e}) = (2.043 \pm 0.018_{\rm stat} \pm 0.012_{\rm syst})\%$ and $\mathcal{B}(D^0\rightarrow K^{*}(892)^-\mu^+\nu_{\mu}) = (1.964 \pm 0.018_{\rm stat} \pm 0.012_{\rm syst})\%$, which are the most precise measurements to date and improves the previous results~\cite{D0Kspiev,2504.10867} by a factor of two. 
The hadronic FF parameters in $D^0\rightarrow K^{*}(892)^-\ell^+\nu_{\ell}$ decays are measured to be $r_{V} = V(0)/A_1(0) = 1.444 \pm 0.026_{\rm stat} \pm 0.010_{\rm syst}$, $r_{2} = A_2(0)/A_1(0) = 0.752 \pm 0.020_{\rm stat} \pm 0.004_{\rm syst}$. Taking into account the $D^0$ lifetime, 
$\tau_{D^0}=410.3\pm1.0~{\rm fs}$, and $|V_{cs}|=0.97435\pm0.00016$~\cite{pdg24}, we obtain $A_1(0)=0.618\pm0.002_{\rm stat} \pm0.004_{\rm syst}$. Furthermore, for the first time, we obtain
$\mathcal{B}(D^0\rightarrow K_2^{*}(1430)^-e^+\nu_e)=(4.00\pm1.22_{\rm stat}\pm0.78_{\rm syst})\times 10^{-5}$, and 
$\mathcal{B}(D^0\rightarrow K_2^{*}(1430)^-\mu^+\nu_{\mu})=(3.85\pm1.17_{\rm stat}\pm0.75_{\rm syst})\times 10^{-5}$, which are consistent with the calculations from the SU(3) flavor symmetry approach~\cite{EPJC84_1110} and the RQM~\cite{prd111_093001} within two standard deviations.

\begin{figure}[tp!]
\begin{center}
   \begin{minipage}[t]{8cm}
   \includegraphics[width=\linewidth]{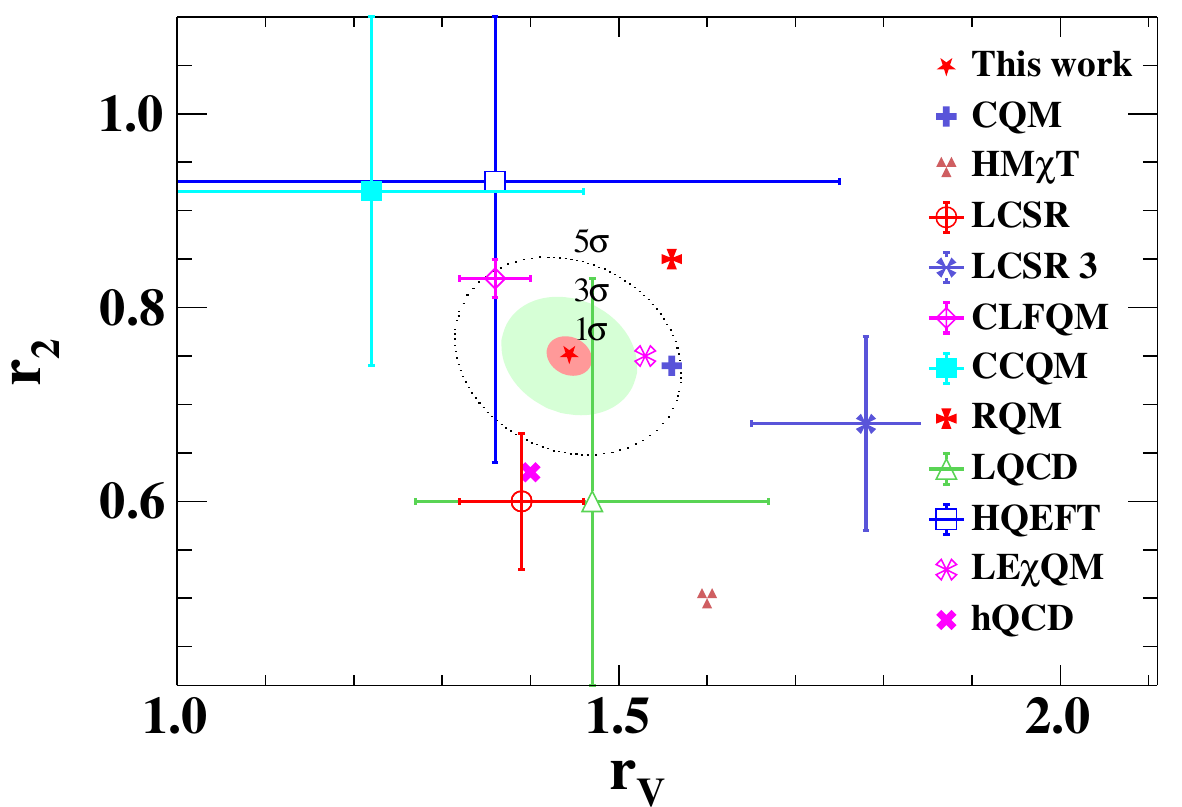}
   \end{minipage}    
   \caption{(Color online)~Comparisons of the measured $r_V$ and $r_2$ in this work with various theoretical calculations from the CQM~\cite{PRD62_014006,PRD96_016017}, HM$\chi$T~\cite{PRD72_034029},  LCSR~\cite{IJMPA21_6125}, LCSR~3~\cite{2505.01329}, CLFQM~\cite{JPG39_025005,EPJC77_587}, CCQM~\cite{FrontPhys14_64401}, RQM~\cite{prd101_013004}, LQCD~\cite{Lattice},HQEFT~\cite{PRD67_014024}, LE$\chi$QM~\cite{PRD89_034013}, hQCD~\cite{prd109_026008}, and the PDG averages~\cite{pdg24}. The correlation coefficient between the measured $r_V$ and $r_2$ is $-0.17$. }      
\label{fig:cmpformfactor}
\end{center}
\end{figure}

The comparisons of $\mathcal{B}(D^0\rightarrow K^*(892)^-e^+\nu_{e})$, $\mathcal{B}(D^0\rightarrow K^*(892)^-\mu^+\nu_{\mu})$, $r_{V}$, and $r_{2}$ between this measurement and theoretical calculations~\cite{Lattice,IJMPA21_6125,PRR2,2505.01329,PRD62_014006,PRD96_016017,PRD67_014024,PRD72_034029,JPG39_025005,EPJC77_587,PRD89_034013,FrontPhys14_64401,prd101_013004,PRD92_054038,prd109_026008} are shown in Figs.~\ref{fig:cmpbf} and~\ref{fig:cmpformfactor}.
Some of the models are ruled out and others are disfavored; 
however, the differing treatments of theoretical uncertainties
(including their omission) makes precise statements difficult in many cases.
Nonetheless, the value of the newly-obtained experimental precision
is evident from these figures.  

Furthermore, using $\mathcal{B}(D^0\rightarrow K^{*}(892)^-e^+\nu_{e})$ and $\mathcal{B}(D^0\rightarrow K^{*}(892)^-\mu^+\nu_{\mu})$ measured in this work
and accounting for correlated uncertainties, we obtain the relative ratios between $\mu$ and $e$ channels: 
\begin{equation}
\mathcal{R}^{\mu/e}_{K^*(892)}=0.961\pm0.012_{\rm stat}\pm0.005_{\rm syst}.
\end{equation}
This ratio are in good agreement with the calculations in Refs.~\cite{IJMPA21_6125,PRD96_016017,EPJC77_587,FrontPhys14_64401,PRD89_034013,prd101_013004}, but disfavor the calculation in Ref.~\cite{PRD92_054038,PRR2} at 95\% C.L.  The results presented in this Letter provide the most powerful tests and constraints on various theoretical calculations especially for QCD theory, and play an important role in understanding
the dynamics of SL decays of the charmed hadrons in the non-perturbative region.

\newpage
\acknowledgments
The BESIII Collaboration thanks the staff of BEPCII (https://cstr.cn/31109.02.BEPC) and the IHEP computing center for their strong support. This work is supported in part by National Key R\&D Program of China under Contracts Nos. 2020YFA0406400, 2020YFA0406300, 2023YFA1606000, 2023YFA1606704; National Natural Science Foundation of China (NSFC) under Contracts Nos. 11635010, 11935015, 11935016, 11935018, 12022510, 12025502, 12035009, 12035013, 12061131003, 12192260, 12192261, 12192262, 12192263, 12192264, 12192265, 12221005, 12225509, 12235017, 12375090, 12475092, 12361141819; the Chinese Academy of Sciences (CAS) Large-Scale Scientific Facility Program; the Strategic Priority Research Program of Chinese Academy of Sciences under Contract No. XDA0480600; CAS under Contract No. YSBR-101; 100 Talents Program of CAS; The Institute of Nuclear and Particle Physics (INPAC) and Shanghai Key Laboratory for Particle Physics and Cosmology; ERC under Contract No. 758462; German Research Foundation DFG under Contract No. FOR5327; Istituto Nazionale di Fisica Nucleare, Italy; Knut and Alice Wallenberg Foundation under Contracts Nos. 2021.0174, 2021.0299; Ministry of Development of Turkey under Contract No. DPT2006K-120470; National Research Foundation of Korea under Contract No. NRF-2022R1A2C1092335; National Science and Technology fund of Mongolia; Polish National Science Centre under Contract No. 2024/53/B/ST2/00975; STFC (United Kingdom); Swedish Research Council under Contract No. 2019.04595 and 2021.04567; U. S. Department of Energy under Contract No. DE-FG02-05ER41374. This paper is also supported by the Fundamental Research Funds for the Central Universities, and the Research Funds of Renmin University of China under Contract No. 24XNKJ05.




\begin{thebibliography}{99}
\bibitem{BFK0} The BFs of $K^0\rightarrow K^0_S$ and $K^0_S\rightarrow \pi^+\pi^-$ are not included in the detection efficiency. 
The $K^0_S$ fraction of a $K^0$ is taken as 50\% and the BF of $K^0_S\rightarrow \pi^+\pi^-$ is taken from PDG~\cite{pdg24}.
\end{thebibliography}
\end{document}